\documentclass[twoside,12pt]{amsart}
\usepackage{amsmath, amssymb, latexsym, amsthm, mathrsfs, color, geometry} 

\long\def\forget#1\forgotten{}
\newcommand{\nc}{\newcommand}
\nc{\x}{\times}
\nc{\Z}{\mathbb{Z}}\nc{\Zp}{\Z_p}\nc{\Zpp}{\Z_{p^2}}
\nc{\bbF}{\mathbb{F}}
\nc{\bpf}{\begin{proof}} \newcommand{\epf}{\end{proof}}
\nc{\op}{\operatorname}
\nc{\mx}[1]{\begin{pmatrix}#1\end{pmatrix}}
\nc{\ord}[1]{{|#1|}} 
\nc{\lcm}{\op{lcm}}
\nc{\my}[1]{\textcolor{red}{#1}}

\newtheorem{thm}{Theorem}
\nc{\bthm}{\begin{thm}} \nc{\ethm}{\end{thm}}
\newtheorem{red}[thm]{Reduction} 
\newtheorem{lem}[thm]{Lemma}
\nc{\blem}{\begin{lem}} \nc{\elem}{\end{lem}}
\newtheorem{cor}[thm]{Corollary}
\nc{\bcor}{\begin{cor}} \nc{\ecor}{\end{cor}}
\nc{\bred}{\begin{red}} \nc{\ered}{\end{red}}
\theoremstyle{definition}
\newtheorem{defn}[thm]{Definition}
\nc{\bdfn}{\begin{defn}} \nc{\edfn}{\end{defn}}

\title[DLP in Bergman's ring]{
The Discrete Logarithm Problem in Bergman's non-representable ring}

\author{Matan Banin}
\author{Boaz Tsaban}
\address{Department of Mathematics, Bar-Ilan University, Ramat Gan 52900, Israel}
\email{baninmmm@gmail.com, tsaban@math.biu.ac.il}
\urladdr{http://www.cs.biu.ac.il/\~{}tsaban}

\begin{document}

\begin{abstract}
Bergman's Ring $E_p$, parameterized by a prime number $p$,
is a ring with $p^5$ elements that cannot be embedded in a ring of matrices over any commutative ring.
This ring was discovered in 1974.
In 2011, Climent, Navarro and Tortosa described an efficient implementation of $E_p$
using simple modular arithmetic, and suggested that this ring may be a useful source
for intractable cryptographic problems.

We present a deterministic polynomial time reduction of the Discrete Logarithm Problem in $E_p$
to the classical Discrete Logarithm Problem in $\Zp$, the $p$-element field.
In particular, the Discrete Logarithm Problem in $E_p$ can be solved, by conventional computers,
in sub-exponential time.
\end{abstract}

\maketitle

\section{introduction}

For Discrete Logarithm based cryptography, it is desirable to find efficiently implementable
groups for which sub-exponential algorithms for the Discrete Logarithm Problem are not available.
Thus far, the only candidates for such groups seem to be (carefully chosen) groups of points
on elliptic curves \cite{Koblitz87, Miller86}.
Groups of invertible matrices over a finite field, proposed in \cite{OSV},
where proved by Menezes and Wu \cite{MW} inadequate for this purpose.
Consequently, any candidate for a platform group for Discrete Logarithm based cryptography
must not be efficiently embeddable in a group of matrices.

In 1974, Bergman proved that the ring $\op{End}(\Zp \times \Zpp)$
of endomorphisms of the group $\Zp \times \Zpp$, where $p$ is a prime parameter,
admits no embedding in any ring of matrices over a commutative ring \cite{Bergman}.
In 2011, Climent, Navarro and Tortosa \cite{CliNaTo11} described an efficient implementation of $E_p$ (reviewed below),
proved that uniformly random elements of $E_p$ are invertible with probability greater than $1-2/p$, and supplied
an efficient way to sample the invertible elements of $E_p$ uniformly at random.
Consequently, they proposed this ring as a potential source for intractable cryptographic problems.
Climent et al.\ proposed a Diffie--Hellman type key exchange protocol over $E_p$, but it was shown
by Kamal and Youssef \cite{KY} not to be related to the Discrete Logarithm Problem,
and to be susceptible to a polynomial time attack.

We consider the Discrete Logarithm Problem in $E_p$.
Since $E_p$ admits no embedding in any ring of matrices over a commutative ring,
the Menezes--Wu reduction attack \cite{MW} is not directly
applicable. We present, however, a
deterministic polynomial time reduction of the Discrete Logarithm Problem in $E_p$
to the classical Discrete Logarithm Problem in $\Zp$, the $p$-element field.
In particular, the Discrete Logarithm Problem in $E_p$ can be solved by conventional computers
in sub-exponential time, and $E_p$ offers no advantage, over $\Zp$, for cryptography based on
the Discrete Logarithm Problem.

\section{Computing discrete logarithms in $\op{End}(\Zp \times \Zpp)$}

Climent, Navarro and Tortosa \cite{CliNaTo11} provide the following faithful representation of
Bergman's Ring. The elements of $E_p$ are the matrices
$$g=\mx{
a & b\\
cp & v+up
}, \quad a,b,c,u,v \in \{0,...,p-1\}.$$
Addition (respectively, multiplication)
is defined by first taking ordinary addition  (respectively, multiplication) over the integers,
and then reducing each element of the first row modulo $p$, and each element of the second row modulo $p^2$.
The ordinary zero and identity integer matrices serve as the additive and multiplicative neutral elements of $E_p$,
respectively.
The element $g$ is invertible in $E_p$ if and only if $a,v \ne 0$.

The group of invertible elements in a ring $R$ is denoted $R^*$.
For an element $g$ in a group, $\ord{g}$ denotes the order of $g$ in that group.

\bdfn
The \emph{Discrete Logarithm Problem} in a ring $R$
is to find $x$ given an element $g\in R^*$ and its power $g^x$, where $x\in \{0,1,\dots,\ord{g}-1\}$.
\edfn

Another version of the Discrete Logarithm Problem asks to find any $\tilde x$ such that $g^{\tilde x}=g^x$. The reductions given below are applicable, with minor changes, to this version as well, but it is known
the two versions are essentially equivalent (see Appendix \ref{equiv}).

By the standard amplification techniques, one can increase the success probability of any
discrete logarithm algorithm with non-negligible success probability to become
arbitrarily close to $1$.
Thus, for simplicity, we may restrict attention to algorithms
that never fail.
For ease of digestion, we present our solution to the Discrete Logarithm Problem in $E_p$ by
starting with the easier cases, and gradually building up. Not all of the easier reductions
are needed for the main ones, but they do contain some of the important ingredients of the main
ones, and may also be of independent interest to some readers.

\subsection{Basic reductions}

\bred\label{computeorder}
Computing the order of an element in $R^*$, using discrete logarithms in $R$.
\ered
\bpf[Details]
For $g\in R^*$, $g^{-1}=g^{\ord{g}-1}$. Thus, $\ord{g}=\log_g(g^{-1})+1$.
\epf

\bred\label{dlprod}
Computing discrete logarithms in a product of rings using discrete logarithms in each ring separately.
\ered
\bpf[Details]
For rings $R,S$, $(R\x S)^*=R^*\x S^*$. Let $(g,h)\in R^*\x S^*$ and $(g,h)^x=(g^x,h^x)$,
where $x\in\{1,\dots,\ord{(g,h)}\}$, be given.
Compute
\begin{eqnarray*}
x \bmod \ord{g} & = & \log_g(g^x);\\
x \bmod \ord{h} & = & \log_h(h^x).
\end{eqnarray*}
Use Reduction \ref{computeorder} to compute $ \ord{g}$ and $\ord{h}$.
Compute, using the Chinese Remainder Algorithm,
$$x \bmod \lcm(\ord{g},\ord{h}) = x \bmod \ord{(g,h)} = x.\qedhere$$
\epf

The \emph{Euler isomorphism} is the function
\begin{eqnarray*}
\Phi_p\colon(\Zp,+)\x(\Zp^*,\cdot) & \to & \Zpp^*\\
(a,b) & \mapsto & (1+ap)\cdot b^p \bmod p^2.
\end{eqnarray*}
The function $\Phi_p$ is easily seen to be an injective homomorphism between groups
of equal cardinality, and thus an isomorphism of groups (cf.\ Paillier \cite{Paillier}
in a slightly more involved context).
The Euler isomorphism can be inverted efficiently:
Given $c\in\Zpp^*$, let $a\in\Zp, b\in\Zp^*$ be such that $c=(1+ap)b^p \bmod p^2$.
Then
$$c = (1+ap)\cdot b^p = 1\cdot b^p = b \pmod p.$$
Compute $b= c \bmod p$, then $b^p \bmod p^2$, then $1+ap=c\cdot (b^p)^{-1} \mod p^2$,
where the inverse is in $\Zpp^*$. Since $1+ap<p^2$, we can subtract $1$ and divide by $p$ to get $a$.

\bred
Computing discrete logarithms in $\Zpp$ using discrete logarithms in $\Zp$.
\ered
\bpf[Details]
Use the Euler isomorphism to transform the problem into a computation of a discrete logarithm in
$(\Zp,+)\x(\Zp^*,\cdot)$. Computing discrete logarithm in $(\Zp,+)$ is trivial. Apply Reduction \ref{dlprod}.
\epf

\subsection{Algebraic lemmata}

\bdfn
$\bar E_p$ is the ring of matrices
$\mx{a&b\\ {pc} & {v}}$, $a,b,c,v\in\{0,1,\dots,p-1\}$,
where addition and multiplication are
carried out over $\Z$, and then entry $(2,1)$ is reduced modulo $p^2$,
and the other three entries are reduced modulo $p$.
\edfn

\blem\label{barhom}
The map
\begin{eqnarray*}
E_p & \to & \bar E_p;\\
\mx{a&b\\ {cp}&{v+up}} & \mapsto & \mx{a&b\\ cp &v}
\end{eqnarray*}
is a ring homomorphism.
\elem
\bpf
Since addition is component-wise, it remains to verify multiplicativity.
Indeed, in $E_p$,
$$
\mx{a_1 & b_1\\ c_1p & v_1+ u_1p}
\mx{a_2 & b_2\\ c_2p & v_2 + u_2p}
=
\mx{a_1 a_2 & a_1b_2 + b_1v_2\\
(c_1a_2 + v_1c_2)p & v_1v_2+(c_1b_2+v_1u_2+u_1v_2)p},
$$
and in $\bar E_p$,
$$
\mx{a_1 & b_1\\ c_1p & v_1}
\mx{a_2 & b_2\\ c_2p & v_2}
=
\mx{a_1 a_2 & a_1b_2 + b_1v_2\\
(c_1a_2 + v_1c_2)p & v_1v_2}.\qedhere
$$
\epf

\blem\label{barpow}
Let $\bar{g}=\mx{a & b\\ cp & v}\in \bar E_p^*$, and let $x$ be a natural number.
Define $d_x\in\Zp$ by
$$d_x=\begin{cases}
\frac{a^x-v^x}{a-v} & a\neq v\\
xa^{x-1} & a = v.
\end{cases}$$
Then
$$\bar{g}^x = \mx{a^x & bd_x\\ cd_xp & v^x}.$$
\elem
\bpf
By induction on $x$. The statement is immediate when $x=1$.
Induction step: If $a\neq v$, then in $\Zp$,
\begin{eqnarray*}
a^x+d_xv & = & a^x+\frac{a^x-v^x}{a-v}\cdot v = \frac{a^x(a-v)+(a^x-v^x)v}{a-v}=\frac{a^{x+1}-v^{x+1}}{a-v}=d_{x+1};\\
ad_x+v^x & = & \frac{a(a^x-v^x)}{a-v}+\frac{(a-v)v^x}{a-v}=\frac{a^{x+1}-v^{x+1}}{a-v}=d_{x+1}.
\end{eqnarray*}
If $a=v$, then
\begin{eqnarray*}
a^x+d_xv & = & a^x+xa^{x-1}v=a^x+xa^{x-1}a=a^x+xa^x=(x+1)a^x=d_{x+1};\\
ad_x+v^x & = & xa^x+a^x=(x+1)a^x=d_{x+1}.
\end{eqnarray*}
Thus, in either case,
$$ \bar{g}^{x+1}=\bar{g}^x\cdot \bar{g}
=\mx{a^x & bd_x\\ cd_xp & v^x} \cdot
\mx{a & b\\ cp & v} =\mx{a^{x+1} & b(a^x+d_xv)\\ c(ad_x+v^x)p & v^{x+1}}
=\mx{a^{x+1} & bd_{x+1}\\ cd_{x+1}p & v^{x+1}}.
$$
\epf

\blem\label{fund}
Let $\bar{g}=\mx{a & b\\ cp & v}\in \bar E_p^*$.
\begin{enumerate}
\item If $a=v$ and at least one of $b,c$ is nonzero, then $\ord{\bar{g}}=p\cdot\ord{a}$.
\item In all other cases ($a\ne v$ or $b=c=0$), $\ord{\bar{g}}=\lcm(\ord{a},\ord{v})$.
\end{enumerate}
\elem
\bpf
Define $d_x$ as in Lemma \ref{barpow}.
By Lemma \ref{barpow},
$$\mx{a^\ord{\bar{g}} & *\\ * & v^\ord{\bar{g}}}=\bar{g}^\ord{\bar{g}}=\mx{1&0\\0&1}.$$
Thus, $\ord{a}$ and $\ord{v}$ divide $\ord{\bar{g}}$, and therefore
so does $\lcm(\ord{a},\ord{v})$.

We consider all possible cases.

If $b=c=0$, then
$$\bar{g}^x=\mx{a^x & 0\\ 0 & v^x}$$
for all $x$, and thus $\ord{\bar{g}}=\lcm(\ord{a},\ord{v})$, as claimed in (2).

Assume, henceforth, that at least one of $b,c$ is nonzero, and let
$$l=\lcm(\ord{a},\ord{v}).$$
If $a\neq v$, then
$$d_l=\frac{a^l-v^l}{a-v} = \frac{1-1}{a-v}=0 \mod p,$$
and thus, by Lemma \ref{barpow},
$\bar{g}^l=I$. Thus, $\ord{\bar{g}}$ divides $l$, which we have seen to divide $\ord{\bar{g}}$.
It follows that $\ord{\bar{g}}=l$, as claimed in (2).

Assume, henceforth, that $a=v$.

Since  $d_p=pa^{p-1} = 0 \mod p$, we have by Lemma \ref{barpow} that
$$\bar{g}^p=\mx{a^p&0\\0&a^p}=\mx{a&0\\0&a}.$$
It follows that $\bar{g}^{p\cdot\ord{a}}=I$. Therefore, $\ord{\bar{g}}$ divides $p\cdot\ord{a}$.
Recall that $\ord{a}$ divides $\ord{\bar{g}}$.
Now, $d_\ord{a}=\ord{a}\cdot a^{\ord{a}-1} \bmod p$. Since $\ord{a}<p$,
$d_\ord{a}\neq 0$. It follows that
$$\bar{g}^\ord{a}=\mx{a^\ord{a}&bd_\ord{a}\\cd_\ord{a}p&a^\ord{a}}\neq \mx{1&0\\0&1},$$
and thus $\ord{\bar{g}}=p\cdot\ord{a}$, as claimed in (1).
\epf

\subsection{The main reductions}

\bred\label{EpBarDL}
Computing discrete logarithms in $\bar E_p$ using discrete logarithms in $\Zp$.
\ered
\bpf[Details]
Let $\bar{g}=\mx{a & b\\ cp & v}\in \bar E_p^*$, and let $x\in\{1,\dots,\ord{\bar{g}}\}$.
By Lemma \ref{barpow},
$$\bar{g}^x = \mx{a^x & bd_x\\ cd_xp & v^x}.$$

If $a\ne v$ or $b=c=0$, then by Lemma \ref{fund}, $\ord{\bar{g}}=\lcm(\ord{a},\ord{v})$.
Compute
\begin{eqnarray*}
x \mod \ord{a} & = & \log_a(a^x);\\
x \mod \ord{v} & = & \log_v(v^x).
\end{eqnarray*}
Since $x<\ord{\bar{g}}$, we can use the Chinese Remainder Algorithm to compute
$x \bmod \lcm(\ord{a},\ord{v})=x$.

Thus, assume that $a=v$ and one of $b,c$ is nonzero.
By Lemma \ref{fund}, $\ord{\bar{g}}=p\cdot\ord{a}$.
Compute
$$x_0:=x \mod \ord{a} = \log_a(a^x).$$
Compute
$$\bar{g}^x\cdot \bar{g}^{-x_0}=\bar{g}^{x-x_0}=\mx{a^{x-x_0} & bd_{x-x_0}\\cd_{x-x_0}p & a^{x-x_0}}=
\mx{1 & bd_{x-x_0}\\cd_{x-x_0}p & 1}.$$
Since $b$ or $c$ is nonzero, we can extract $d_{x-x_0} \bmod p$.
Compute
$$d_{x-x_0}\cdot a=(x-x_0)a^{x-x_0}=x-x_0\mod p.$$
As $x-x_0\le x<\ord{\bar{g}}=p\cdot\ord{a}$, we can use the Chinese Remainder Algorithm to compute
$$x-x_0 \bmod \lcm(p,\ord{a}) = x-x_0 \bmod p\cdot\ord{a} = x-x_0.$$
Add $x_0$ to obtain $x$.
\epf

\bred
Computing discrete logarithms in $E_p$ using discrete logarithms in $\Zp$.
\ered
\bpf[Details]
Let $g=\mx{a & b\\ cp & v+up}\in E_p^*$, and let $x\in\{1,\dots,\ord{g}\}$.
Take $\bar{g} = \mx{a & b\\ cp & v}\in \bar E_p^*$.
Use Lemma \ref{fund} and Reduction  \ref{computeorder}
to compute $\ord{\bar{g}}$.
By Lemma \ref{barhom}, $\ord{\bar{g}}$ divides $\ord{g}$.
As $\bar{g}^{\ord{\bar{g}}}=I$ is the image of $g^{\ord{\bar{g}}}$ under the homomorphism
of Lemma \ref{barhom}, we have that
$$g^{\ord{\bar{g}}}=\mx{1 & 0\\ 0 & 1+sp}$$
for some $s\in\{0,\dots,p-1\}$.
Using Reduction \ref{EpBarDL}, compute
$$x_0:=\log_{\bar{g}}(\bar{g}^x)=x \bmod \ord{\bar{g}}.$$
If $s=0$ then $\ord{g}=\ord{\bar{g}}$, and thus
$x_0:=\log_{\bar{g}}(\bar{g}^x)=\log_g(g^x)=x$,
and we are done.
If $s\neq 0$, let $q=(x-x_0)/ \ord{\bar{g}}$.
Since the order of $1+sp$ in $\Zpp$ is $p$ (in $\Zpp$, $(1+sp)^e=1+esp$ for all $e$),
the order of $g^{\ord{\bar{g}}}$ is $p$, and thus
$\ord{g}=\ord{\bar{g}}\cdot p$.
Thus, $q\le x/\ord{\bar{g}}<\ord{g}/\ord{\bar{g}}=p$.
Compute
$$g^xg^{-x_0}=g^{x-x_0}=(g^{\ord{\bar{g}}})^q=\mx{1 & 0\\ 0 & 1+sp}^q=
\mx{1 & 0\\ 0 & (1+sp)^q}=\mx{1 & 0\\ 0 & 1+sqp}.$$
Compute $sq \bmod p =((1+sqp)-1)/p$.
In $\Zp$, multiply by $s^{-1}$ to obtain $q\bmod p=q$.
Multiply by $\ord{\bar{g}}$ to get $x-x_0$, and add $x_0$.
\epf

\section{Summing up: Code}

Following is a self-explanatory code (in Magma \cite{Magma}) of our main
reductions. This code shows, in a concise manner,
that the number of computations of discrete logarithms
in $\Zp$ needed to compute discrete logarithms in Bergman's Ring $E_p$ is \emph{at most 2}.
For completeness, we provide, in Appendix \ref{basicode}, the basic routines.

\begin{verbatim}
F := GaloisField(p);
Z := IntegerRing();
I :=  ScalarMatrix(2, 1); //identity matrix

function EpBarOrder(g) //Lemma 9
    a := F!(g[1,1]);
    v := F!(g[2,2]);
    if (a ne v) or (IsZero(g[1,2]) and IsZero(g[2,1])) then
        order := Lcm(Order(a),Order(v));
    else
        order := p*Order(a);
    end if;
    return order;
end function;

function EpBarLog(g,h) //Reduction 10
    a := F!(g[1,1]);
    b := F!(g[1,2]);
    c := F!(g[2,1] div p);
    v := F!(g[2,2]);
    x0 := Log(a,F!(h[1,1]));
    if (a ne v) or (IsZero(b) and IsZero(c)) then
        xv := Log(v,F!(h[2,2]));
        x := ChineseRemainderTheorem([x0,xv], [Order(a),Order(v)]);
    else
        ginv := EpBarInverse(g);
        f := EpBarPower(ginv,x0);
        f := EpBarProd(h,f);
        if IsZero(c) then
            d := b^-1 * F!(f[1,2]);
        else
            d := c^-1 * F!(f[2,1] div p);
        end if;
        delta := Z!(d*a);
        truedelta := ChineseRemainderTheorem([0,delta],[Order(a),p]);
        x := truedelta+x0;
    end if;
    return x;
end function;

function EpLog(g,h) //Reduction 11
    gbar := Bar(g); hbar := Bar(h);
    gbarorder := EpBarOrder(gbar);
    x0 := EpBarLog(gbar,hbar);

    f := EpPower(g,gbarorder);
    s := (f[2,2]-1) div p;

    if IsZero(s) then
        x := x0;
    else
        ginv := EpInverse(g);

        f := EpPower(ginv,x0);
        f := EpProd(h,f);
        n := (f[2,2]-1) div p;
        q := (F!s)^-1*F!n;
        x := gbarorder*(Z!q)+x0;
    end if;
    return x;
end function;
\end{verbatim}

We have tested these routines extensively:
For random primes of size $4,8,16,32,64$, and $128$ bits, and thousands of random pairs $g,h=g^x$,
\verb|EpLog(g,h)| always returned $x$.

\appendix

\section{Elementary routines}\label{basicode}

To remove any potential ambiguity, and help readers interested in reproducing our experiments,
we provide here the basic routines for arithmetic in Bergman's Ring $E_p$.

\begin{verbatim}
function EpProd(A, B) //integer matrices
    C := A*B;
    C[1,1] mod:= p;
    C[1,2] mod:= p;
    C[2,1] mod:= p^2;
    C[2,2] mod:= p^2;
    return C;
end function;

function Bar(g)
    h := g;
    h[2,2] mod:= p;
    return h;
end function;

function EpBarProd(A, B) //integer matrices
    return Bar(EpProd(A,B));
end function;

function EpInvertibleEpMatrix()
    g := ZeroMatrix(Z, 2, 2);
    g[1,1] := Random([1..p-1]);
    g[1,2] := Random([0..p-1]);
    g[2,1] := p*Random([0..p-1]);
    g[2,2] := Random([1..p-1])+p*Random([1..p-1]);
    return g;
end function;

function EpPower(g, n) //square and multiply
     result := I;
     while not IsZero(n) do
        if ((n mod 2) eq 1) then
            result := EpProd(result, g);
            n -:= 1;
        end if;
        g := EpProd(g, g);
        n div:= 2;
     end while;
     return result;
end function;

function EpBarPower(g, n)
    return Bar(EpPower(g, n));
end function;

function EpInverse(g)
    a := F!(g[1,1]);
    b := F!(g[1,2]);
    c := F!(g[2,1] div p);
    u := F!(g[2,2] div p);
    v := F!(g[2,2]);

    ginv := ZeroMatrix(Z,2,2);
    ginv[1,1] := Z!(a^-1);
    ginv[1,2] := Z!(-a^-1*b*v^-1);
    ginv[2,1] := p*Z!(-v^-1*c*a^-1);
    ginv[2,2] := Z!(v^-1)+
            p*Z!(c*a^-1*b*v^-2-u*v^-2-(F!(Z!v*Z!(v^-1) div p)*v^-1));
    return ginv;
end function;

function EpBarInverse(g)
    return Bar(EpInverse(g));
end function;
\end{verbatim}

\section{Equivalence of Discrete Logarithm Problems}\label{equiv}

The result in this appendix should be well known to experts, but since we are not aware of any reference
for it, we include it for completeness.
Consider the following two versions of the Discrete Logarithm Problem in a prescribed finite group $G$.
We assume that $|G|$, or a polynomial upper bound $K$ on $|G|$, is known. We do not assume that
$G$ is cyclic.

\begin{description}
\item[DLP1]
Find $x$, given an element $g\in G$ and its power $g^x$, where $x\in \{0,1,\dots,\ord{g}-1\}$.
\item[DLP2]
Given an element $g\in G$ and its power $g^x$, find $\tilde x$ with $g^{\tilde x}=g^x$.
\end{description}

DLP1 is harder than DLP2: A DLP1 oracle returns $\tilde x:=x \bmod \ord{g}$ on input $g,g^x$.
On the other hand, DLP2 is probabilistically harder than DLP1: It suffices to show how
$\ord{g}$ can be computed using a DLP2 oracle. Indeed, for a large enough (but polynomial) number
of random elements $r\in \{K,K+1,\dots,M\}$ where $M\gg K$ is fixed,
let $\tilde r$ be the output of DLP2 on $(g,g^r)$. Then $\ord{g}$ divides
all numbers $(r-\tilde r)\bmod g$, and the greatest common divisor of these numbers is $\ord{g}$, except for a
negligible probability.


\begin{thebibliography}{99}

\bibitem{Bergman}
G. Bergman, \emph{Examples in PI ring theory},
Israel Journal of Mathematics \textbf{18} (1974), 257�-277.

\bibitem{Magma}
W. Bosma, J. Cannon, and C. Playoust, \emph{The Magma algebra system, I: The user language},
Journal of Symbolic Computation \textbf{24} (1997), 235--265.

\bibitem{CliNaTo11}
J. Climent, P. Navarro, L. Tortosa,
\emph{On the arithmetic of the endomorphisms ring $\operatorname{End}(\Zp \times \Zpp)$},
Applicable Algebra in Engineering, Communication and Computing \textbf{22} (2011), 91--108.

\bibitem{KY}
A. Kamal, A. Youssef,
\emph{Cryptanalysis of a key exchange protocol based on the endomorphisms ring $\operatorname{End}(\Zp \times \Zpp)$},
Applicable Algebra in Engineering, Communication and Computing, to appear.

\bibitem{Koblitz87}
N. Koblitz, \emph{Elliptic curve cryptosystems}, Mathematics of Computation \textbf{48}  (1987), 203--209.


\bibitem{MW}
A. Menezes, Y. Wu, \emph{The discrete logarithm problem in $\op{GL}(n,q)$},
Ars Combinatoria, 47 (1998), 23--32.

\bibitem{Miller86}
V. Miller, \emph{Uses of elliptic curves in cryptography}, in: Advances in Cryptology--Proceedings of
Crypto '85. Lecture Notes in Computer Science \textbf{218} (1986), 417--426.

\bibitem{OSV}
R. Odoni, R. Sanders, V. Varadharajan, \emph{Public key distribution in matrix rings},
Electronic Letters \textbf{20} (1984), 386--387.

\bibitem{Paillier}
P. Paillier, \emph{Public-key cryptosystems based on composite degree residuosity classes},
in: J. Stern, ed., Advances in Cryptology -- EUROCRYPT'99,
Lecture Notes in Computer Science \textbf{1592} (1999), 223--238.

\end{thebibliography}
\end{document}